\documentclass[letterpaper]{ptephy_v1}
\usepackage{amsmath}
\usepackage{graphicx}
\usepackage{color}
\usepackage{caption}
\usepackage{subcaption}

\newcommand{\beq}{\begin{equation}}
\newcommand{\eeq}{\end{equation}\noindent}
\newcommand{\bal}{\begin{align}}
\newcommand{\eal}{\end{align}}
\newcommand{\nn}{\nonumber}
\newcommand{\TR}{\mbox{Tr}}
\newcommand{\mDU}{\mathcal{D}U}

\graphicspath{{figs/}}

\begin{document}

\title{Sign problem in finite density lattice QCD}


\author[1,2,*]{V.~A.~Goy}
\affil{School of Natural Sciences, Far Eastern Federal University, Sukhanova 8, 690950 Vladivostok, Russia \email{vovagoy@gmail.com}}

\author[2,3]{V.~Bornyakov}
\affil{School of Biomedicine, Far Eastern Federal University, Sukhanova 8, 690950 Vladivostok, Russia}

\author[2]{D.~Boyda}
\affil{ITEP, B. Cheremushkinskaya 25, Moscow, 117218 Russia}

\author[2]{A.~Molochkov}

\affil{Research Center for Nuclear Physics (RCNP), Osaka University, Ibaraki, Osaka 567-0047, Japan}
\author[2,4,5]{A.~Nakamura}
\affil{Theoretical Research Division, Nishina Center, RIKEN, Wako 351-0198, Japan}

\author[2]{A.~Nikolaev}

\author[2,6]{V.~Zakharov}
\affil{Moscow Inst Phys $\&$ Technol, Dolgoprudny, Moscow Region 141700, Russia}

\begin{abstract}%
The canonical approach, which was developed for solving the sign
problem, may suffer from a new type of sign problem.  
In the canonical approach, the grand partition function is
written as a fugacity expansion:
$Z_G(\mu,T) = \sum_n Z_C(n,T) \xi^n$, where 
$\xi=\exp(\mu/T)$ is the fugacity, and $Z_C(n,T)$ are given
as averages over a Monte Carlo update, $\langle z_n\rangle$.  
We show that the complex phase of $z_n$ is proportional to $n$
at each Monte Carlo step. Although $\langle z_n\rangle$ take
real positive values, the values of $z_n$ fluctuate rapidly
when $n$ is large, especially in the confinement phase, 
which gives a limit on $n$. We discuss possible remedies for
this problem. 
\end{abstract}

\subjectindex{D31,D34}

\maketitle

\noindent
{\it 1. Sign Problem and the Canonical Approach} \ \ 
A lattice QCD simulation is a first-principles calculation, and this
makes it possible to study the quark/hadron world using a non-perturbative approach.
The basic formula is the Feynman path integral form of the grand
partition function:
\beq
Z_G(\mu,T) = \TR\, e^{-(\hat{H}-\mu \hat{N})/T} 
= \int \mDU (\det\Delta(\mu))^{N_f} e^{-S_G},
\eeq
where $\mu$ is the chemical potential, $T$ is
the temperature, $\hat{H}$ is the Hamiltonian, $\hat{N}$ is the number operator, $\det\Delta(\mu)$ is the fermion determinant,
and $S_G$ is the gluon kinetic energy.
In this paper, we consider the two-flavor case: $N_f=2$.  

To explore the finite density QCD, we consider finite
$\mu$ regions.  However, when $\mu$ takes a nonzero real value,
the fermion determinant becomes a complex number. 
This is problematic, because in the Monte Carlo simulations,
we generate the gluon fields with the probability
\beq
P = (\det\Delta(\mu))^{N_f} e^{-S_G}/Z,
\eeq
and if the fermion determinant is complex, we are in trouble.
In principle, we may write $\det\Delta=|\det\Delta|\exp(i \phi)$,
perform the Monte Carlo update with $|\det\Delta|$, and 
push the phase $\exp(i \phi)$ into an observable. 
However, the fermion determinant may have the form 
$\det\Delta=\exp(-f V/T)$, and the phase fluctuation $\mbox{Im} f V/T$
becomes large as the volume becomes large, so this does not work
in practice.

Recently, the canonical approach, or the fugacity expansion,
has attracted much attention as a candidate
for solving the sign problem
\cite{Borici:2004bq,
Alexandru:2005ix,
deForcrand:2006ec,
nagata2010wilson,
Alexandru:2010yb,
li2010finite,
Li:2011ee,
Nagata:2012tc,
Danzer:2012vw,
Gattringer:2014hra}.
In the canonical approach, the grand canonical partition function is
expressed as a fugacity expansion:
\beq
Z_G(\mu,T) = \sum_{n=-\infty}^{+\infty} Z_C(n,T) \xi^n,
\label{Eq:FugacityExpansion}
\eeq
where
$\xi = \exp(\mu/T)$ is the fugacity.
Both $Z_G$ and $Z_c$ are functions of the volume,$V$, which we abbreviate 
in the arguments.

The inverse transformation is
\beq
Z_C(n,T) = \int_0^{2\pi}\frac{d\phi}{2\pi} e^{-in\phi} Z_G(\xi=e^{i\phi},T).
\label{Eq:InverseFugacityExpansion}
\eeq

In Eq.\ (\ref{Eq:FugacityExpansion}), 
the canonical partition functions,
$Z_C$, do {\it not} depend on $\mu$,
and Eq.\ (\ref{Eq:FugacityExpansion}) works for real, 
imaginary, and even complex $\mu$.
When the chemical potential is pure imaginary, $\mu=i\mu_I$,
the fermion determinant is real, and in those regions,
we can construct $Z_C$ from $Z_G$.  
After determining $Z_C$ in this way, we can
study real physical $\mu$ regions using formula 
(\ref{Eq:FugacityExpansion}).

\bigskip
\noindent
{\it 2. Calculation of $Z_C(n)$}\ \ 
In order to obtain the canonical partition functions, $Z_C(n,T)$,
first we expand the fermion determinant:
\beq
(\det\Delta(\mu))^{N_f} = \sum_n z_n(U) \xi^n.
\eeq
Then,
\beq
\frac{Z_G(\mu)}{Z_G(\mu_0)} 
= \frac{1}{Z_G(\mu_0)} \int \mDU \left(\frac{\det\Delta(\mu)}{\det\Delta(\mu_0)}
\right)^{N_f}
\det\Delta(\mu_0)^{N_f} e^{-S_G}
=\left\langle
\frac{\sum_n z_n(U) \xi^n}{\det\Delta(\mu_0)^{N_f}}
\right
\rangle_0 .
\eeq
Here, $\langle\cdots\rangle_0$ is an expectation value at $\mu_0$.
One can assign any pure imaginary value to $\mu_0$.

There are various ways to obtain $z_n$:
\begin{enumerate}
\item
Direct evaluation of $\det\Delta$ \cite{nagata2010wilson}.
\item
Fourier transformation \cite{hasenfratz1992canonical}.
\item
Winding number expansion 
\cite{
collaboration2015beating,
nakamura2015study}
\end{enumerate}
In this letter, we employ method (3), in which the fermion determinant
$\det\Delta$ is expanded as a series of the hopping parameter $\kappa$,
and the diagrams are  classified and packed with respect to the fugacity power:
\begin{align}
(\det\Delta)^{N_f} &= \left(\det(I-\kappa Q(\mu))\right)^{N_f}
\nn
\noindent \\
&= \exp \left( N_f \TR \log (I-\kappa Q) \right)
=\exp \left(-N_f \TR\sum_{m=1}^{M_{max}} \frac{\kappa^m}{m} Q^m \right)
\label{Eq:WindNum1}
\noindent \\
&\rightarrow \exp\left(\sum_{k=-K_{max}}^{K_{max}} W_k \xi^k \right)
\label{Eq:WindNum2}
\noindent \\
&\sim \sum_{-N_{max}}^{N_{max}} z_n \xi^n ,
\label{Eq:WindNum3}
\end{align}
where $N_{max}=2N_c N_x N_y N_z$.
An algorithm that describes how to obtain Eq.\ (\ref{Eq:WindNum2}) from Eq.\ (\ref{Eq:WindNum1}) 
is given as
``Algorithm 1: Winding Numbers via Hopping Parameter Expansion''
in Ref.\ \cite{collaboration2015beating}.

\begin{figure}[ht]
  \centering
  \begin{subfigure}{.5\textwidth}
    \centering
    \includegraphics[width=0.9\textwidth]{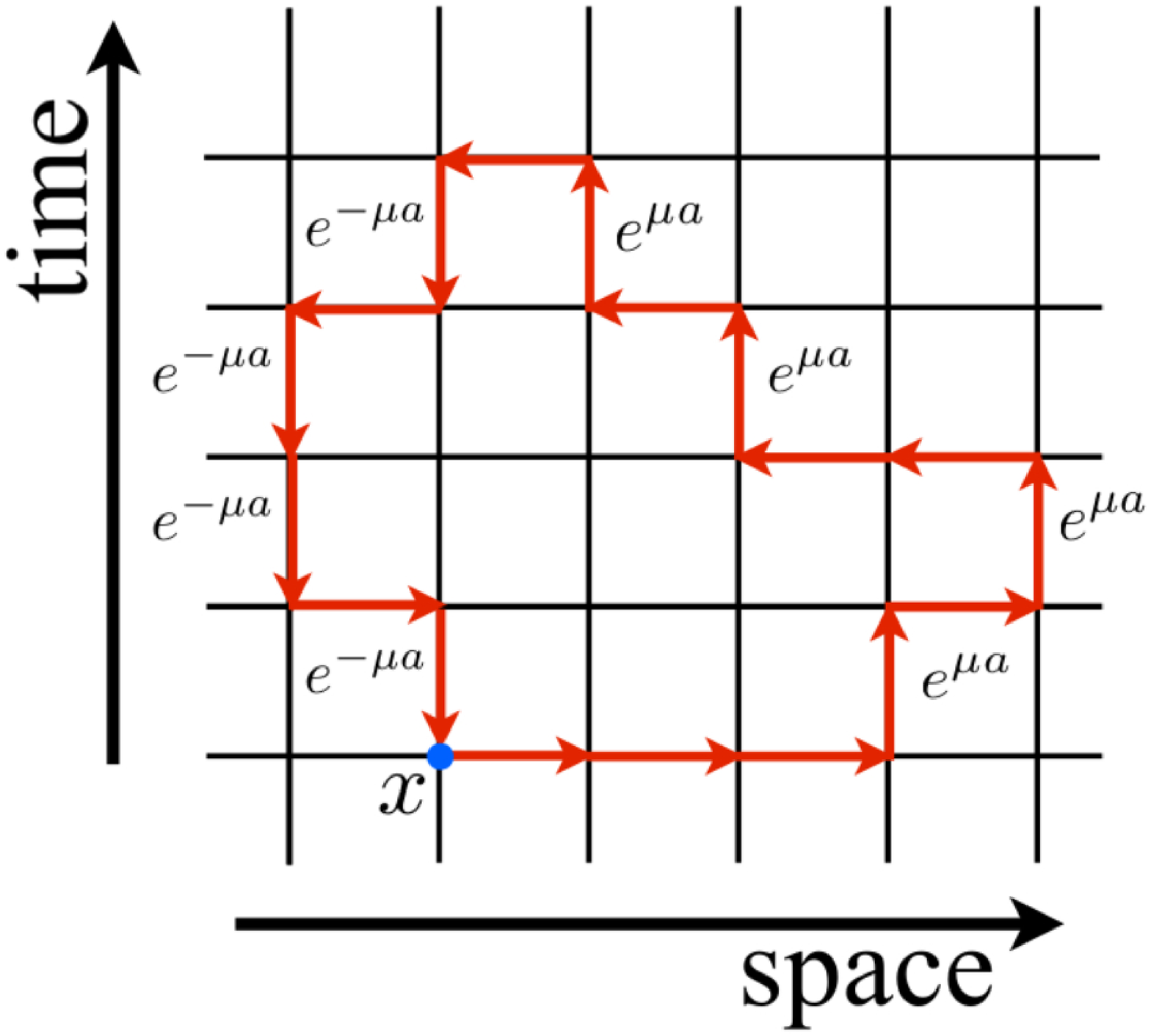}
    \label{Fig:WindingN0}
  \end{subfigure}%
  \begin{subfigure}{.5\textwidth}
    \centering
    \includegraphics[width=0.9\textwidth]{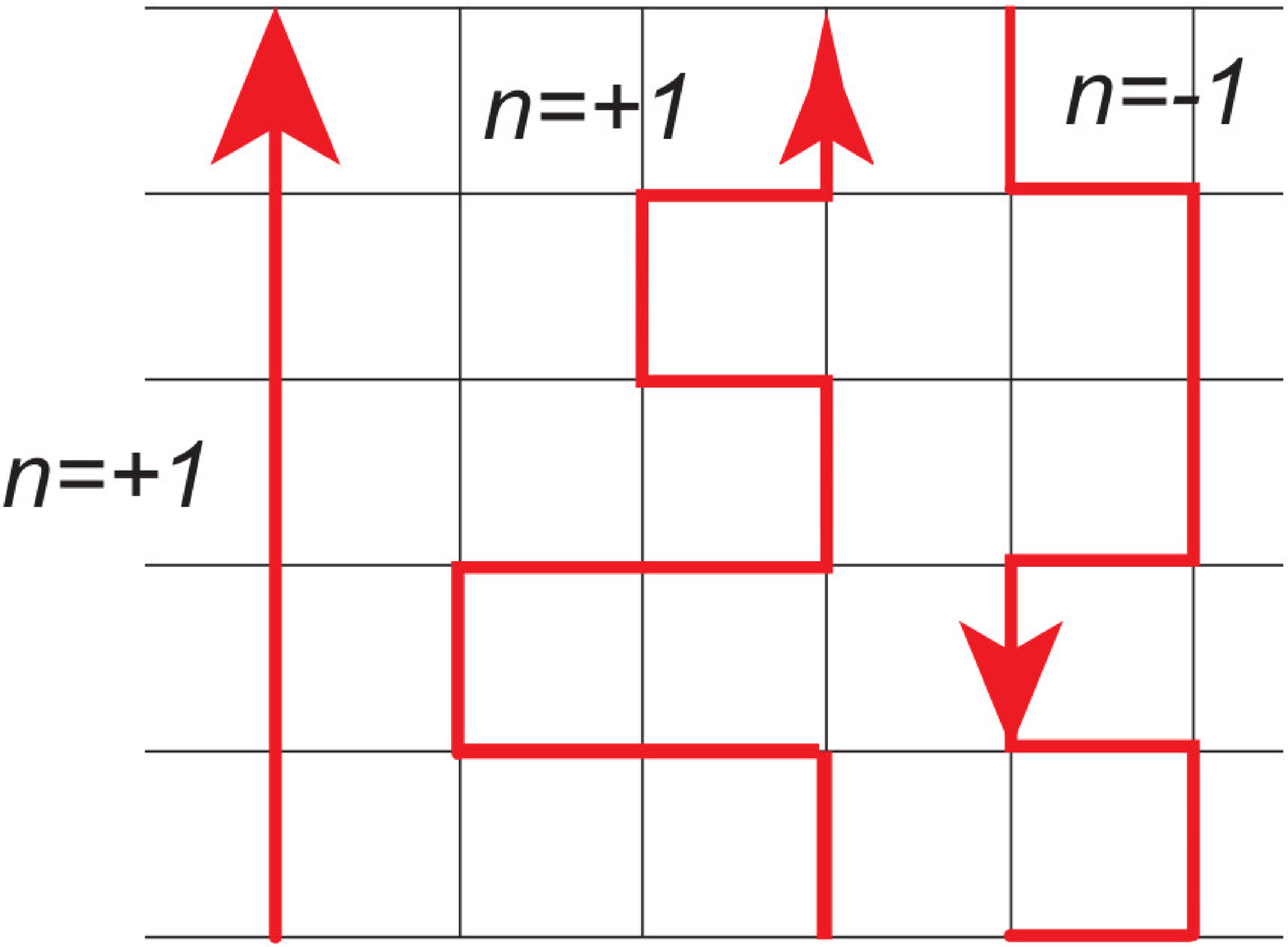}
    \label{Fig:WindingN1}
  \end{subfigure}
  \caption{Schematic of the winding diagrams: Left, $n=0$; Right, $n=\pm 1$.}
  \label{Fig:WindingN}
\end{figure}

\begin{figure}[ht]
  \centering
  \includegraphics[width=.8\textwidth]{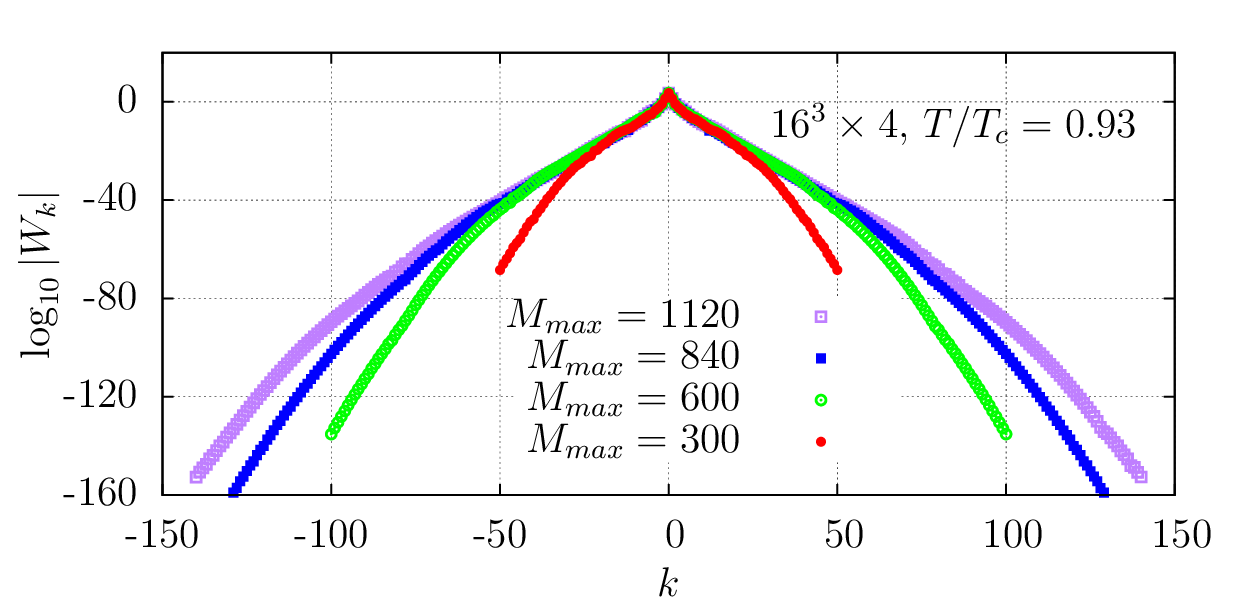}
  \caption{Winding numbers, $W_k$ in Eq.\ (\ref{Eq:WindNum2})
  as a function of $k$ for $M_{max}=$ $300$, $600$, $840$, and $1120$.
  Here, $M_{max}$ is the limit of the sum in Eq.\ (\ref{Eq:WindNum1}).
  }
  \label{Fig:WnNmaxDep}
\end{figure}

In Fig.\ \ref{Fig:WindingN}, we show some of the winding diagrams
of Eq.\ (\ref{Eq:WindNum1}), which contribute $W_0$, $W_1$, and $W_{-1}$.
In Fig.\ \ref{Fig:WnNmaxDep}, the magnitude of the winding numbers $|W_k|$
is shown as a function of $k$, for $M_{max}=$ $300$, $600$, $840$, and $1120$
in the confinement temperature region ($T/T_c=0.93$).

\bigskip
\noindent
{\it 3. Result}\ \ 
The lattice QCD simulations reported here were performed at the Far Eastern Federal University on Vostok-1, which consists of
10 nodes (2 $\times$ Intel E5-2680\_v2, 64 GB RAM; 2 $\times$  Nvidia Tesla K40X Kepler).
Its LINPACK performance is 23.52 TFlops.
In our code, the clover Dirac operator performance was 76.9 GFLOPS 
(53.4\% of the peak on the GTX 980).

The lattice size $N_x N_y N_z\times N_t$ is $16^3\times 4$, and $\beta$ and the hopping parameter
are chosen to ensure $m_\pi/m_\rho= 0.80$
\cite{Ejiri:2009hq}.

\begin{figure}[ht]
  \centering
  \includegraphics[width=0.8\linewidth]{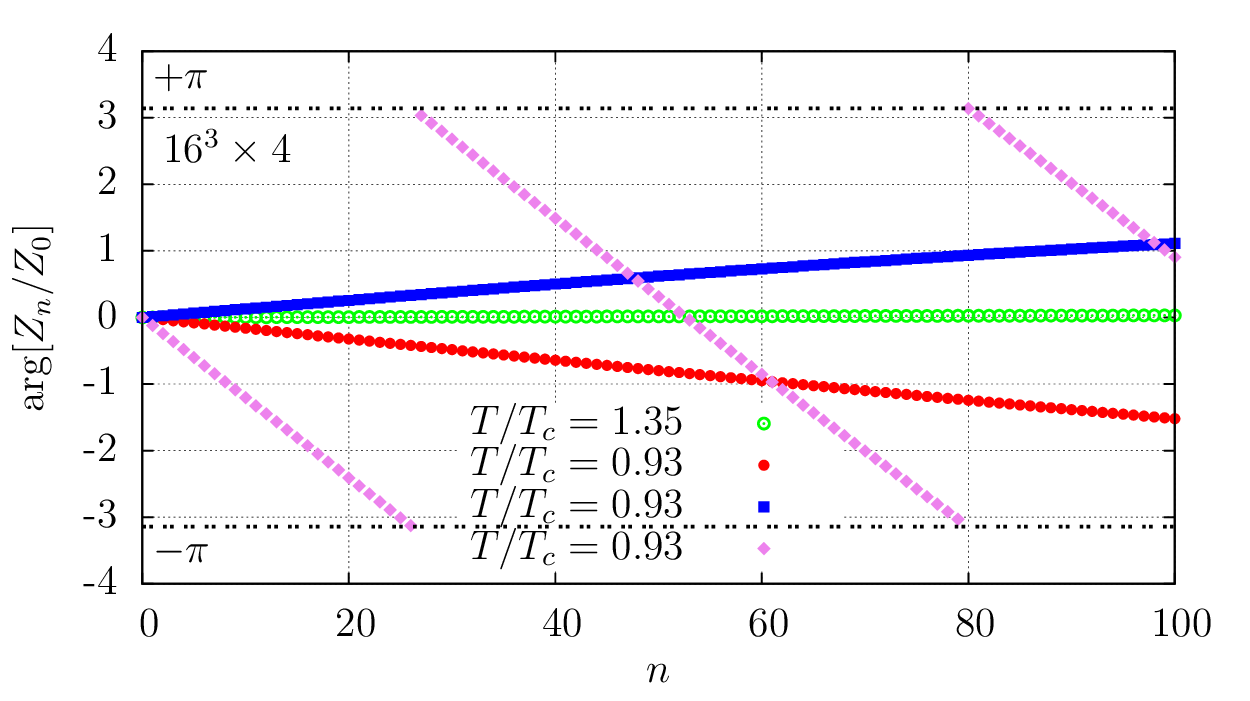}
  \vspace*{-2ex}
  \caption{
  Phase of $z_n$ for several configurations in the deconfinement and 
  confinement regions. 
  }
  \label{Fig:ZnPhase}
\end{figure}

In Fig.\ \ref{Fig:ZnPhase}, we show the complex phase of $z_n$ for
several configurations.  The first observation is that they
are approximately proportional to $n$.
In Ref.\ \cite{li2010finite}, the authors presented the relationship between
the winding number expansion and the canonical partition functions, as follows: 
\begin{align}
\det\Delta(\mu) 
 &= \exp \left( A_0 + 
\sum_{k>0} [ e^{i k\phi}W_k + e^{-i k\phi}W_k^\dagger
]
\right)
\nn \\
&= \exp \left( A_0 + \sum_k A_k\cos(k\phi + \delta_k) \right).
\label{Eq:WindNumExp0}
\end{align}
Here, 
\beq
A_0 \equiv W_0,\quad A_k \equiv 2|W_k| \quad\mbox{and}\quad
\delta_k \equiv arg(W_k),
\eeq
and we use the relation
\beq
W_{-k} = W_k^\dagger.
\eeq
 
Then, we obtain the Fourier transform of Eq.\ (\ref{Eq:WindNumExp0}) 
to get $Z_C$ in Eq.\ (\ref{Eq:InverseFugacityExpansion})\footnote{For $N_f$ flavors, we make replacement $A_n \to N_f A_n$.},
\beq
\int_0^{2\pi} \frac{d\phi}{2\pi} e^{-in\phi} e^{A_0 + A_1\cos(\phi+\delta_1)
+ A_2\cos(2\phi+\delta_2) \cdots}.
\label{Eq:FTofWn}
\eeq

Using the integral representation of the modified Bessel function,
\beq
I_n(z) = \frac{(-1)^n}{2\pi} \int_0^{2\pi} e^{z\cos t} e^{-int} dt,
\eeq
the lowest order of Eq. (\ref{Eq:FTofWn}) reads
\begin{align}
\int_0^{2\pi} \frac{d\phi}{2\pi} e^{-in\phi} e^{A_0 + A_1\cos(\phi+\delta_1)}
&=e^{A_0} \int_{\delta_1}^{2\pi+\delta_1} \frac{d\phi'}{2\pi} 
e^{-in(\phi'-\delta_1)} e^{A_1\cos\phi'}
\nn \\
&=e^{A_0+in\delta_1} \int_{\delta_1}^{2\pi+\delta_1} \frac{d\phi'}{2\pi} 
e^{-in\phi'} e^{A_1\cos\phi'}
\nn \\
&=e^{A_0+in\delta_1} \int_{0}^{2\pi} \frac{d\phi'}{2\pi} 
e^{-in\phi'} e^{A_1\cos\phi'}
=e^{A_0 + in\delta_1} I_n(A_1).
\label{Eq:ModBesselInt}
\end{align}

This is proportional to $z_n$ in each configuration.
In this lowest order, we see that $z_n$ has a phase coming from
$e^{i n\delta_1}$, which is proportional to $n$.

The phase of $z_n$ is approximately proportional to $n$.
In the following, we parametrize the phase as 
$n\delta$.
In the deconfinement phase, the slope is small, namely, $z_n$ are 
nearly real, while in the confinement phase, the slope is large, 
sometimes crossing $\pm\pi$.

\begin{figure}[ht]
\centering
\includegraphics[width=0.8\linewidth]{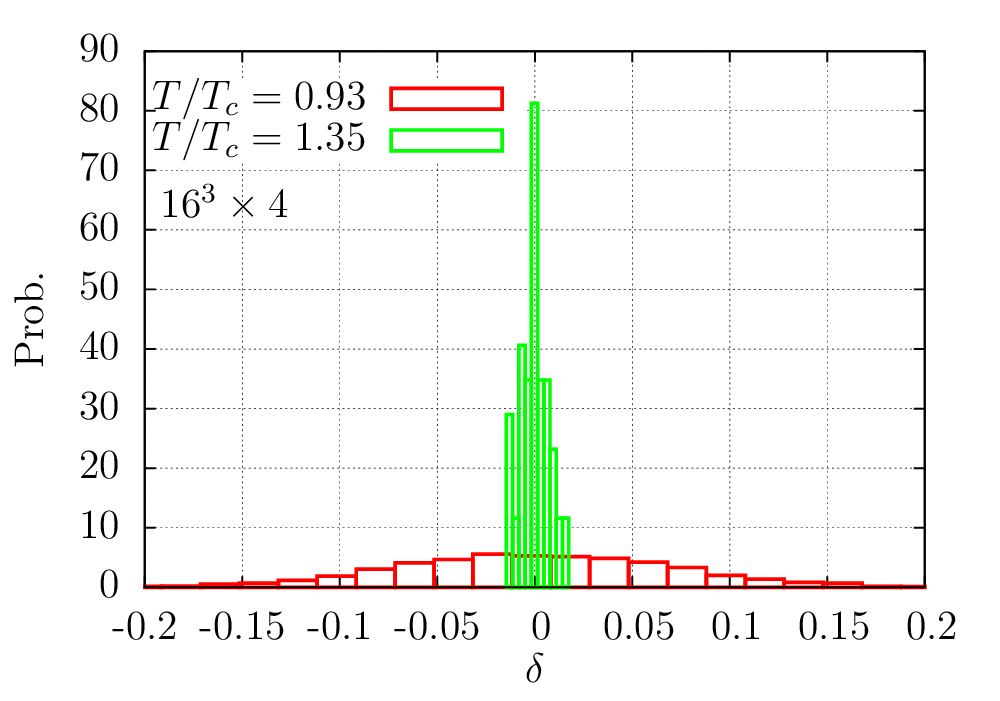}
\vspace*{-2ex}
\caption{
   Distribution of the complex phase of $z_n$ ($\arg z_n = n\delta$) for the confinement and
  deconfinement phases.
}
\label{Fig:HistPhase}
\end{figure}

The distribution of $\delta$ is shown in Fig.\ \ref{Fig:HistPhase},
and the  very different behavior in the confinement and
deconfinement phases is apparent.
Above $T_c$, the phase of $z_n$ is almost zero, while
below $T_c$, the phase fluctuates significantly.
For example, if $\delta=0.1$, $n\delta>\pi$ for $n>31$.
In other words, for large $n$, the real part of $z_n$ fluctuates
between positive and negative values in the confinement phase.
After averaging many configurations, the average of $z_n$ should
be real positive.  But when $n$ is large, we suffer from a ``sign problem''. 

\begin{figure}[ht]
\centering
\includegraphics[width=0.8\linewidth]{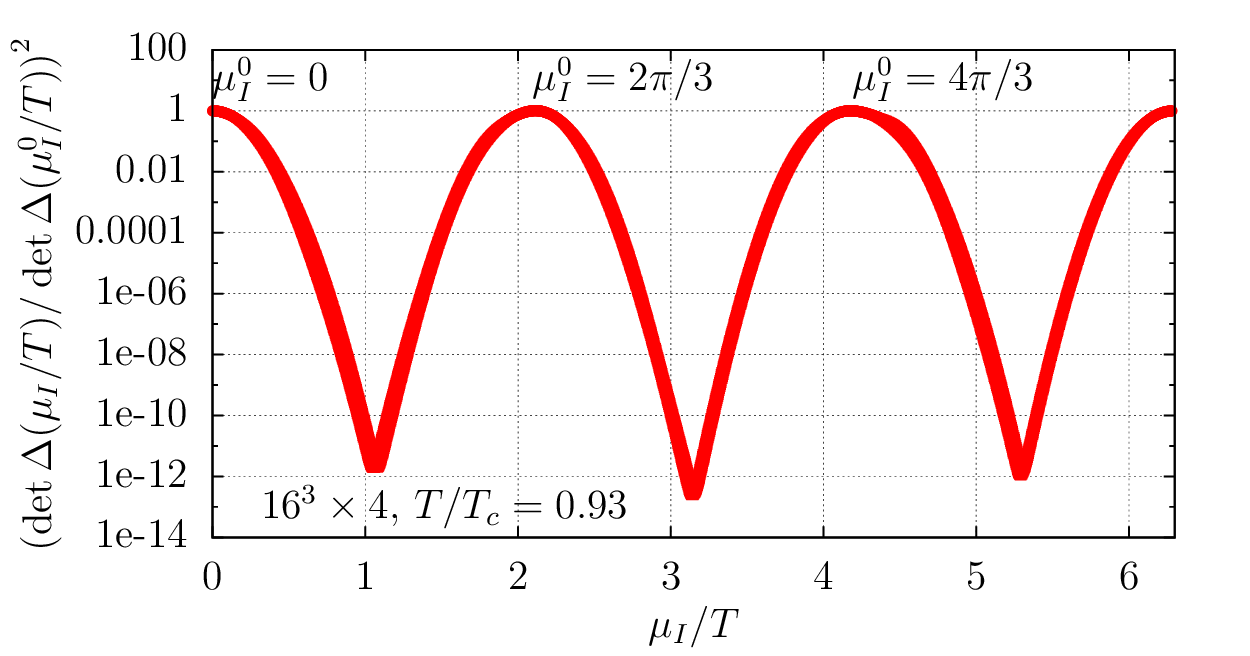}
\caption{
   $\det\Delta(\mu_I/T)/\det\Delta(\mu_I^0/T)$ as a function of
  $\mu_I/T$. The result is a sum of three evaluations at
  $\mu_I^0/T=0$, $2\pi/3$, and $4\pi/3$.
}
\label{Fig:Triality}
\end{figure}

In Fig.\ \ref{Fig:Triality}, we show the behavior of the fermion determinant 
in the regions in which the chemical potential is pure imaginary.
The data are evaluated by the reweighting method at 
$\mu_I^0/T=0,\, 2\pi/3$, and $4\pi/3$,
in order to recover Roberge-Weiss symmetry.

\bigskip
\noindent
{\it 4. Conclusion}\ \ 
In this letter, we reported that there is a hidden sign 
problem in the canonical approach, namely, $z_n$ having a complex phase.
This was observed in Refs.\ \cite{SuzukiLat2016} and
\cite{collaboration2015beating}.
We have confirmed it and found that it produces positive/negative
cancellation in the confinement phase.
We studied the distribution of the phase, $\delta$, as shown 
in Fig.\ \ref{Fig:HistPhase}. This provides an approximate upper bound on
$|n|$ of $Z_n$: $|n|<\pi/\langle |\delta|\rangle$.

The arguments from Eqs.\ (\ref{Eq:WindNumExp0}) to (\ref{Eq:ModBesselInt})
tell us that the phase of the winding number $W_n$ determines
the phase of $z_n$.
Although each $W_n$ contributes $z_l$, the largest effect comes from
$W_{\pm 1}$.   
These consist of many diagrams, but the ones with the largest contribution are
the Polyakov lines, $L$ and $L^\dagger$.
Therefore, the phase of the Polyakov lines produces the phase of $z_n$.
Values of $L$ and $L^\dagger$ scatter around the real axis,
and at lower temperatures, $|L|$ is small, which results in
the large phase. 

It is known that, in the grand canonical approach, 
the complex Polyakov line contributes the phase of the determinant.
In the canonical approach, we can pinpoint the origin of the
sign problem.
In order to overcome the sign problem,
it is important to pursue the following:
\begin{enumerate}
\item
Study the lattice size and the quark mass dependence.
In particular, we must investigate whether this new sign problem increases or decreases in severity as the lattice volume increases.
\item
Find a solution to reduce this sign problem.
The canonical approach can be combined with the Lefschetz thimbles
method or other practical methods \cite{cristoforetti2014efficient},
\cite{alexandru2016monte}.
\end{enumerate}

\section*{Acknowledgment}
\quad
This work was completed thanks to support of the RSF grant 15-12-20008.
One of the authors (A.~Nakamura) is partially supported by
JSPS KAKENHI Grant Numbers 26610072 and 15H03663.
The calculations were performed on Vostok-1 at FEFU.
We thank all members of the Zn collaboration, especially
Asobu Suzuki and Yusuke Taniguchi, for useful discussions.

\bibliographystyle{ptephy}
\bibliography{FDQCD}

\end{document}